\documentclass[12pt]{article} 
\usepackage{dina4p} 
\usepackage{epsfig}  
\def\lapprox{\lower .7ex\hbox{$\;\stackrel{\textstyle <}{\sim}\;$}}
\def\gapprox{\lower .7ex\hbox{$\;\stackrel{\textstyle >}{\sim}\;$}}

\def\lsim{\mathrel{\rlap{\lower4pt\hbox{\hskip1pt$\sim$}}
    \raise1pt\hbox{$<$}}}                
\def\gsim{\mathrel{\rlap{\lower4pt\hbox{\hskip1pt$\sim$}}
    \raise1pt\hbox{$>$}}}                
\begin{document}
\begin{flushright}
TTP99-36
\end{flushright}
\vspace*{6mm}
\begin{center}  \begin{Large} \begin{bf}
Summary: Spin Physics\\
  \end{bf}  \end{Large}
  \vspace*{5mm}
  \begin{large}
T.~Gehrmann
  \end{large}
\end{center}
Institut f\"ur Theoretische Teilchenphysik,
Universit\"at Karlsruhe, D-76128 Karlsruhe, Germany

\begin{quotation}
\noindent
{\bf Abstract:}
The spin physics parallel sessions at this workshop made a critical
review of the physics potential of future experiments on polarized
nucleons, with an emphasis on the potential impact of polarized
electron-proton collisions at HERA. A summary of the results  
and discussions from these sessions is presented in this article. 
\end{quotation}

\section{Introduction}
A series of polarized lepton-nucleon scattering 
experiments at CERN, SLAC and DESY has considerably improved our
knowledge on the spin structure of the nucleon over the past few years. 
These experiments were however largely restricted to the inclusive
structure function $g_1(x,Q^2)$, probing a particular combination of 
quark polarizations in the nucleon. As a consequence, our picture of the 
nucleon spin structure is still far from complete. Several future
experiments are now attempting to resolve remaining open
issues. The spin physics sessions of this workshop made a critical
review of the physics potential of these future experiments, 
with particular emphasis on the potential impact of 
polarized electron-proton collisions at HERA. 

With several new and ongoing spin experiments, much information on
the proton spin structure will become available in the next few years:
complementary measurements at HERMES, COMPASS and RHIC will yield first
information on the gluon contribution to the proton spin, combining
future results from these experiments with existing structure function
data, it will moreover be possible to separate quark and antiquark
contributions to the proton spin and to carry out a flavour
decomposition. These experiments, as well as future measurements at
Jefferson Laboratory will also give first insight into potential higher
twist contributions to the proton spin.   
The kinematical scope of these experiments is however limited, and they
will still leave many questions unanswered. 

Given the numerous new insights HERA has provided into the unpolarized
proton structure in the past, the option of polarizing its beams to study 
proton spin structure appears very tempting. Polarization of the HERA
electron beam comes naturally due to the Sokolov-Ternov effect,
and is already used for the HERMES experiment. Polarizing the proton
beam is a much more complicated task, and other sessions of this
workshop were devoted to the machine aspects connected with this
project. 
The physics prospects of a polarized HERA collider 
were investigated for the first time in a working 
group of the 1995/96 ``Future Physics at HERA'' workshop~\cite{heraspin}.
The most important observables identified in this 
working group were the polarized structure function $g_1(x,Q^2)$, 
polarized weak structure functions, 
dijet production in polarized DIS and polarized photoproduction of jets. 
The working group established 
the measurability of all these observables, given an integrated 
luminosity of at least 200~pb$^{-1}$.  Encouraged by these results, a 
follow-on workshop ``Physics with Polarized Protons at HERA'' was
organized in 1997~\cite{spin97,revspin}, where more elaborate studies of the
measurability of different aspects of the proton spin structure at HERA 
were carried out. For present workshop, many of these analyses have
been further refined, and a number of new processes, like 
deeply virtual compton scattering or spin
asymmetries in diffraction and in leptoquark production, 
have been investigated in detail 
 for the first time. A new aspect to spin physics at HERA is 
also the possibility of colliding the HERA proton beam with the
polarized electron 
beam of a linear electron-positron collider, whose construction is 
currently considered at DESY. Several studies at this workshop have
illustrated the physics prospects of this project.

\section{Current and future experiments}
The HERMES collaboration is operating a polarized
fixed target spectrometer with final state hadron identification
in the HERA electron beam. In addition to measurements of individual
polarized quark distributions from semi-inclusive
asymmetries~\cite{hermeshad}, this experiment has recently provided a
first glimpse on the polarized gluon distribution~\cite{hermesglue} by 
studying charged hadron production at high transverse
momentum~\cite{bravar}. In the near future, HERMES will study a 
variety of inclusive and semi-inclusive observables in polarized
electron-nucleon scattering. In particular the semi-inclusive 
measurements at HERMES will yield new information on angular
lepton-hadron correlations~\cite{teryaev} and on transverse momentum
dependence (Collins effect) in fragmentation 
processes~\cite{murgia,efremov}.

Running at lower electron energy than HERMES, the polarized deep
inelastic scattering programme at Jefferson Laboratory~\cite{meziani} is 
aiming to study spin structure functions at large $x$ and/or low $Q^2$. 
This kinematic region is of particular interest for the determination of 
higher twist contributions to polarized structure functions, testing 
integral relations~\cite{blumlein}, lattice
calculations~\cite{schierholz} and bounds on asymmetries~\cite{teryaev2}.

The COMPASS experiment~\cite{kunne}, 
which is currently under construction at CERN, will start measuring
inclusive and semi-inclusive observables in polarized lepton-nucleon
scattering next year. With a lepton beam energy significantly above
HERMES, the COMPASS experiment will be able to use charm production as a 
probe of the polarized gluon distribution. COMPASS will also carry out a 
broad programme of inclusive as well as semi-inclusive measurements on 
longitudinally and transversely polarized targets.

A new domain for spin physics will open up with the
commissioning of the Relativistic Heavy Ion Collider (RHIC) at BNL two
years from now. RHIC can be operated with polarized proton beams,
allowing to study polarized proton-proton collisions at
$\sqrt{s}=200\ldots 500$~GeV with two multi-purpose collider detectors. 
The RHIC spin physics programme~\cite{bunce} covers a wide variety of
processes probing different aspects of the nucleon spin structure; first 
detailed simulations~\cite{martin} are now ongoing and have been
reported to the workshop. Polarized quark distributions for individual
flavours can be accessed at RHIC from massive gauge boson production,
simulations indicate that this  measurement should be relatively
unproblematic~\cite{bunce}. The polarized gluon distribution could be
measured at RHIC
from prompt photon or jet production, both channels are however 
not free from experimental as well as theoretical problems. In
particular the direct photon channel has been subject of detailed
simulation~\cite{sowinski}, demonstrating the experimental feasibility 
of measuring this observable. The theoretical interpretation of direct
photon data is however problematic already in the unpolarized case,
since the experimental isolation cuts, which are applied to define the
photon, can hardly be matched in the theoretical calculation. The
extraction of the gluon distribution from these data is therefore not
free from ambiguities, direct photon data are therefore no longer used
in unpolarized fits. The problem of photon isolation will be the same in 
polarized studies, thus casting doubt on the reliability of this
measurement. Alternative photon isolation
criteria~\cite{morgan,frixgam} could allow a better matching of
experiment and theory. Using the 
isolated photon definition proposed in~\cite{frixgam}, 
a first case study for RHIC was carried out
recently~\cite{fv}. 
Jet production at RHIC as a probe of the polarized  
gluon distribution has not yet been investigated in great detail, this
process has the advantage of having large production rates, but suffers
from the numerous competing partonic subprocesses. 

Data from HERMES, Jefferson Laboratory, COMPASS and RHIC will largely
extend our knowledge on several aspects of the nucleon spin structure
over the next few years. These measurements are however limited in their 
kinematical reach, and they will still leave many questions unanswered. 
All three polarized lepton-nucleon scattering experiments are working on 
fixed targets, and are therefore not able to access the behaviour of 
polarized structure functions at small $x$. Also, photoproduction at
these experiments will largely be dominated by direct photon-nucleon
interactions, thus not allowing studies of the spin structure of the
photon. Measurements at the RHIC collider will also cover only the
behaviour of the polarized parton distributions at large and medium $x$, 
which might in particular be problematic for an accurate determination
of the first moment of the polarized gluon distribution.
In unpolarized proton structure, similar aspects are of
high interest, and could first be studied experimentally at the HERA
electron-proton collider. 

\section{Prospects of Spin Physics at HERA}
The operation of  HERA with polarized proton and electron beams  would 
allow to study a wide variety of observables in 
polarized electron--proton collisions at $\sqrt s=300$~GeV. The physics
prospects of this project were first studied in a working group of the
1995/96 workshop ``Future Physics at HERA''~\cite{heraspin} and a
follow-on workshop ``Physics with Polarized Protons at
HERA''~\cite{spin97,revspin} in 1997. 
These workshops identified a number of key observables, which
allow to probe several aspects of the proton and photon spin structure
inaccessible at other experiments. These encouraging physics prospects
also triggered much effort on machine studies for operating HERA with a
polarized proton beam. At the present workshop, latest developments on
machine aspects~\cite{chao} and physics prospects were shown.  
Concerning the physics studies, much effort has gone into 
optimization and refinement of the measurement of key observables as
well as into studies of potential new channels. The most prominent
new results are summarized in the following. 

\subsection{Spin Physics at small $x$ and large $Q^2$}
The unpolarized electron-proton programme at HERA 
has extended the kinematical domain of deep inelastic scattering towards 
small values of the Bjorken scaling variable $x$ and large values of 
the invariant momentum transfer $Q^2$. Both limits probe the structure
of the proton and the dynamics of QCD interactions at a previously
unexplored level and give rise to a multitude of new phenomena, which
have been studied in detail by the two HERA collider experiments over
the past years. 

The kinematical reach of HERA towards large $Q^2$ allows perturbative
QCD evolution of structure functions to be tested over a long range in
$Q^2$, thus allowing for example for an 
indirect determination of the unpolarized gluon distribution from 
$F_2(x,Q^2)$. Likewise, operation of HERA with polarized protons would
allow measurements of $g_1(x,Q^2)$ well beyond the kinematical reach of
present fixed target experiments, thus providing important constraints
to QCD fits and allowing for an indirect extraction of the polarized
gluon distribution from QCD evolution of $g_1(x,Q^2)$. The prospects of
this measurement have been investigated in detail already at previous
workshops~\cite{heraspin,spin97,deroecketal}, some updates have now
made. Concerning detector effects on the extraction of $g_1(x,Q^2)$
from the measured asymmetry, new studies of uncertainties induced 
by bin migration have been carried out~\cite{aidala}, showing in
particular that bin migration effects can be reduced by using the hadron 
method for the reconstruction of the event kinematics.

To study the impact of HERA 
structure function data on a determination of 
polarized parton distributions from QCD fits, projected HERA data points 
have been included in global fits~\cite{deroecketal}. A recent
improvement to these studies is the inclusion of projected charged
current data, which will allow for a more precise determination of the quark
contribution to the proton spin~\cite{lichtenstadt}. 

\begin{figure}[t]
\begin{center}
~ \epsfig{file=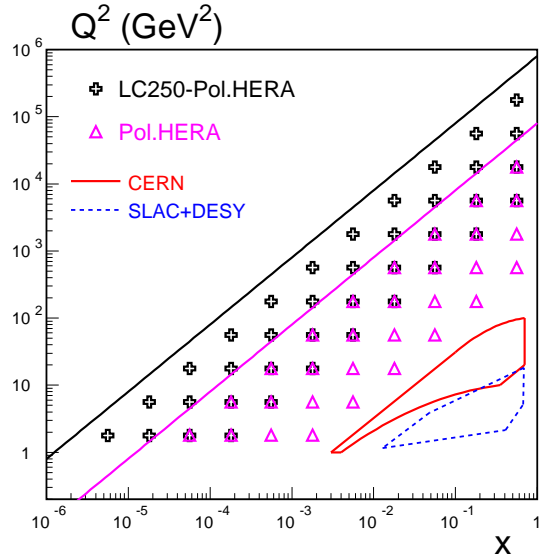,width=7cm}
\end{center}
\caption{Kinematical region accessible with 
linear collider electron beam on HERA
  proton beam, compared to kinematics of present experiments and HERA.}
\label{fig:teslahera}
\end{figure}
A new aspect to measurements of $g_1(x,Q^2)$ at HERA would be the
possibility of colliding the HERA proton beam with the electron beam of
a future linear collider, 
whose construction is currently considered at
DESY. First studies on the kinematical reach of the option were
presented~\cite{deshpande}. Already a linear collider with electron beam 
energy of 250~GeV would enlarge the kinematical reach of HERA by about
an order of magnitude towards large $Q^2$ and small $x$, see
Fig.~\ref{fig:teslahera}.

The behaviour of polarized structure functions at small $x$ is currently 
not known from experiment. 
Present $g_1(x,Q^2)$ data from the SMC collaboration~\cite{smcsmx}
 reach down to 
$x=0.0008$, corresponding however to very small values of
$Q^2=0.2~$GeV$^2$. The
SMC data have motivated several attempts of a theoretical
interpretation~\cite{smallx,badelek}, which are however yet inconclusive. 

In the small-$x$ domain, one would ultimately expect the ln~$Q^2$
resummation of QCD to break down, since terms proportional to ln~$x$
should become of equal importance, thus requiring a reordering of the
perturbative series. HERA measurements of the unpolarized structure
function $F_2(x,Q^2)$ at small $x$ are however still consistent with 
the DGLAP evolution equations, based on ln~$Q^2$ resummation. In this
structure function, one finds the most singular terms at small
$x$ to be of the form $\alpha_s^n\ln^n x$. In the polarized structure
function $g_1(x,Q^2)$, even more singular terms of the form 
$\alpha_s^n\ln^{2n} x$ are present~\cite{ber}, 
resulting in a stronger enhancement at small $x$.  At this workshop, 
a first attempt has been presented~\cite{ziaja1} to combine small-$x$ 
resummation with the usual DGLAP evolution into a unified
evolution equation. As a result, one observes a significant enhancement
of the magnitude of $g_1$ at small $x$, compared to DGLAP evolution
only. A similar enhancement is also expected in $g_2(x,Q^2)$ at small
$x$~\cite{ryskin}; a measurement of this structure function seems
however to be unlikely at HERA. 

\begin{figure}[t]
\begin{center}
~ \epsfig{file=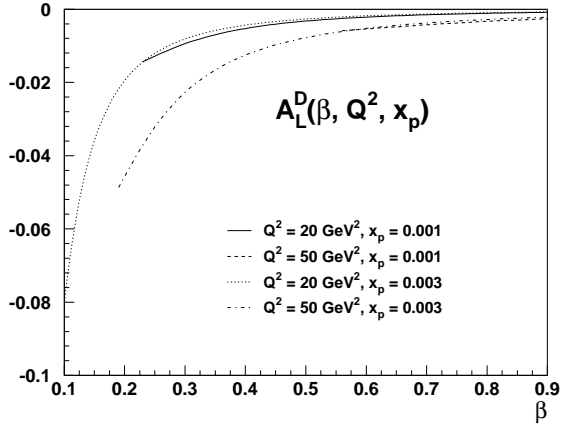,angle=-90,width=7.5cm}
\end{center}
\caption{Expected perturbative 
  spin asymmetry in diffractive deep inelastic
  scattering at HERA in the leading ln~$Q^2$ approximation.}
\label{fig:diff}
\end{figure}
About 10\% of unpolarized deep inelastic scattering events at HERA 
yield a final state containing a diffractively scattered proton
surrounded by a large rapidity gap. Since its first observation 
at HERA several years ago, diffraction in deep inelastic scattering has
triggered a lot of theoretical effort towards its understanding. At
present, it is still fair to say that an unambiguous interpretation of
this phenomenon could not yet be achieved, since both perturbative as
well as non-perturbative effects are expected to contribute with 
comparable magnitude. Theoretical studies on diffraction in 
polarized deep inelastic scattering were presented at this
workshop. Estimations of non-perturbative contributions 
based on regge theory~\cite{mana} predict a  
diffractive asymmetry not exceeding $10^{-4}$, while perturbative
diffractive exchanges~\cite{gehrmann} give rise to asymmetries of the order 
$10^{-2}$, as can been seen in Fig.~\ref{fig:diff}.
 The ratio of perturbative to non-perturbative contributions 
to diffraction appears therefore to be more favourable in the polarized
than in the unpolarized case, in particular due to 
terms of the form $\alpha_s^n\ln^{2n} x$, which are present only
in the polarized, but not in the unpolarized diffractive amplitude. 
Studies of particular final states in polarized diffractive scattering
are also ongoing, large asymmetries are predicted~\cite{golo} for example
in diffractive charm production.

\subsection{Jet production in deep inelastic scattering}
The most promising direct probe of the polarized gluon distribution $\Delta
g(x,Q^2)$ at HERA is the measurement of asymmetries in dijet production
in polarized deep inelastic scattering. Already the past workshops have
proven~\cite{dgpast}
that this observable allows for a determination of  $\Delta
g(x,Q^2)$ down to $x=0.003$, almost an order of magnitude in $x$ below
the reach of measurements at RHIC. Since current QCD fits predict an important 
contribution to the first moment of $\Delta g(x,Q^2)$ from this region,
this measurement appears to mandatory for a determination of the gluon
contribution to the proton spin. 

\begin{figure}[t]
\begin{center}
~ \epsfig{file=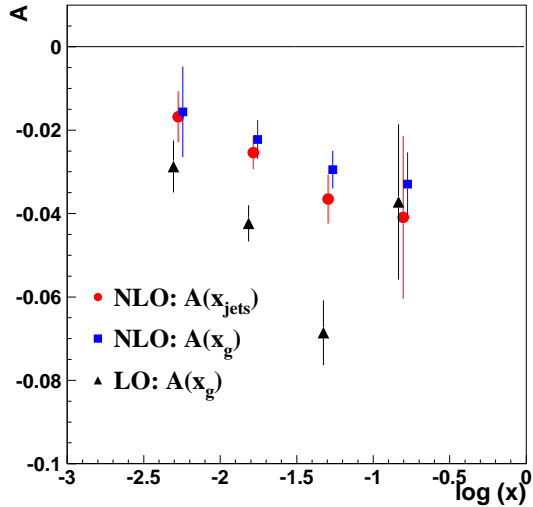,width=7cm}
\end{center}
\caption{Expected spin asymmetry in deep inelastic 
jet production. Cuts according
  to~\protect{\cite{dgnew}}.
  Errors correspond to
  integrated luminosity of 200 pb$^{-1}$ and 70\% polarization of
  electron and proton beams.}
\label{fig:jets}
\end{figure}

Earlier analyses of dijet production 
in polarized deep inelastic scattering have now been extended
considerably~\cite{dgnew} with the incorporation of next-to-leading
order corrections to the parton level subprocesses~\cite{willfahrt}.
It could be shown that the gluon induced scattering process remains 
dominant source of dijet events also at NLO. 
Moreover, the non-trivial correlation between the 
proton energy fraction $x_g$ of the gluon contributing to the scattering
process and the reconstructed $x_{jets}$ from the final state is
now understood at NLO, thus allowing for a consistent extraction of the
polarized gluon distribution at this order. The expected 
jet production asymmetry at next-to-leading order is displayed in
Fig.~\ref{fig:jets}. 

The impact of dijet data on QCD fits of polarized parton distributions
has been investigated in detail~\cite{deroecketal,lichtenstadt}. 
It could be shown
that these data are in particular essential to constrain the 
shape of the gluon distribution. 

The kinematical reach of the dijet
measurement would also be extended by almost another order of magnitude
in $x$ for the option of colliding the electron beam from 
a linear collider with the HERA proton beam~\cite{deshpande}. 
 
As an alternative to the dijet measurement, it has been suggested to use 
current-target correlations to determine the polarized gluon
distribution~\cite{chekanov}. This process has the advantage that 
no requirement on jets in the final state is made, thus including all 
inclusive deep inelastic scattering events in the analysis. The
extraction of $\Delta g(x,Q^2)$ from this process is however far less clean
than from the dijet measurement, large systematic
uncertainties must be expected. 

Jet production in the forward proton region has been suggested as a probe of 
QCD enhancement effects at small $x$ in the unpolarized
case. Measurements of this observable turned out to be substantially
above the expectations obtained by resumming only ln~$Q^2$ terms, thus
indicating the presence of large ln~$x$ in this process. First studies
of small-$x$ resummation effects in polarized forward jet production 
were presented at the workshop~\cite{ziaja2}, indicating that small-$x$ 
resummation results in a characteristic increase of the cross section
for this observable, which could therefore be used as a tool 
to study small-$x$ dynamics in polarized scattering.

\subsection{Polarized Photoproduction}
Cross sections in electron-proton collisions become largest, if 
the virtuality of the  photon mediating the interaction is small. 
In this photoproduction limit, one can approximate the 
electron-proton cross section as a product of a photon flux factor and 
an interaction cross section of a real photon with the proton. 
Many unpolarized photoproduction reactions are presently 
measured at HERA, and their study has continuously improved our knowledge
on proton and photon structure as well as our understanding of the 
transition between real and virtual photons over the last years.

Polarized photoproduction processes have already been in the focus of
the past two workshops~\cite{phoprodold}, where jet photoproduction was
identified as the most promising channel for a determination of both the 
polarized gluon distribution in the proton and the polarized parton
distributions in the photon. 
Jet production in the 
 photon direction
originates mainly from photon-gluon fusion
processes, and thus reflects the gluon polarization in the proton. The 
situation is more involved in the  proton direction, where 
most events are induced by the yet unknown
resolved partonic content of the polarized photon. Given the polarized 
parton distributions in the proton to be known from other sources, jet 
photoproduction in the proton direction can be used to 
determine the polarized parton distributions in the resolved photon.
Next-to-leading order QCD corrections to polarized photoproduction of
jets have recently been calculated~\cite{frixione}, resulting in larger 
perturbative stability of the predictions and enabling a consistent
extraction of parton distributions at this order. 

Knowledge on parton distributions in the polarized photon is essential
to exploit the full physics potential of electron-photon and
photon-photon collisions at a future linear electron-positron collider,
which is foreseen to operate with polarized beams. At present, it seems
that polarized photoproduction of dijets at HERA is the unique process
to determine these distributions. Given that jet production originates
from several different partonic subprocesses, unfolding of all 
polarized quark and  gluon distributions in the photon from dijet
measurements is a very involved, if not impossible task. In the
unpolarized case, one successfully applies the so-called effective
parton approximation~\cite{maxwell}. In this approximation, one replaces 
the sum over different partonic subprocesses, each weighted with a
different parton distribution, by one universal subprocess, weighted
with an effective distribution. This effective distribution is a linear
combination of parton distributions, where each distribution is weighted 
with the approximate importance of the corresponding subprocess. This 
approximation has now been extended to the polarized
case~\cite{stratmann}, where it will help to quantify the expected
precision of a polarized photon structure measurement at HERA.

Polarized photoproduction of charmed 
quarks has been discarded as competitive probe of 
the polarized gluon distribution already at previous
workshops~\cite{phoprodold} due to the expected low charm tagging
efficiency. The production of open charm and of bound states of charmed
quarks has however received renewed interest at this
workshop as testing ground for models for charm production~\cite{sudoh} 
and charm fragmentation~\cite{arestov}.

\subsection{Deeply virtual compton scattering}
The exclusive process $\gamma^*p\to \gamma p$ (deeply virtual compton
scattering, DVCS) can be used to probe aspects of the proton
structure beyond the observables usually studied in inclusive and 
semi-inclusive deep inelastic scattering~\cite{dvcs}. 
It particular, this reaction
allows a measurement of non-forward (skewed) parton distributions, which 
are universal quantities expected to arise in diffractive and exclusive
scattering processes. A direct measurement of the cross section for
DVCS is however not possible at HERA, 
since this process is concealed by a large 
background from photon bremsstrahlung in ordinary electron-proton 
scattering. By measuring single spin asymmetries in exclusive photon
production in the scattering of polarized electrons off unpolarized
protons, it is however possible to access the interference term of the 
DVCS amplitude with the DIS amplitude. First numerical studies of this
process have been presented at the workshop~\cite{strikman}. These are
 indicating
large single spin asymmetries, which will allow a determination of 
skewed parton distributions at HERA, once both collider experiments are
operating with longitudinal electron polarization. Adding proton
polarization would enable a measurement of polarized skewed parton
distributions as well.

\subsection{Searches for new phenomena} 
\begin{figure}[t]
\begin{center}
~ \epsfig{file=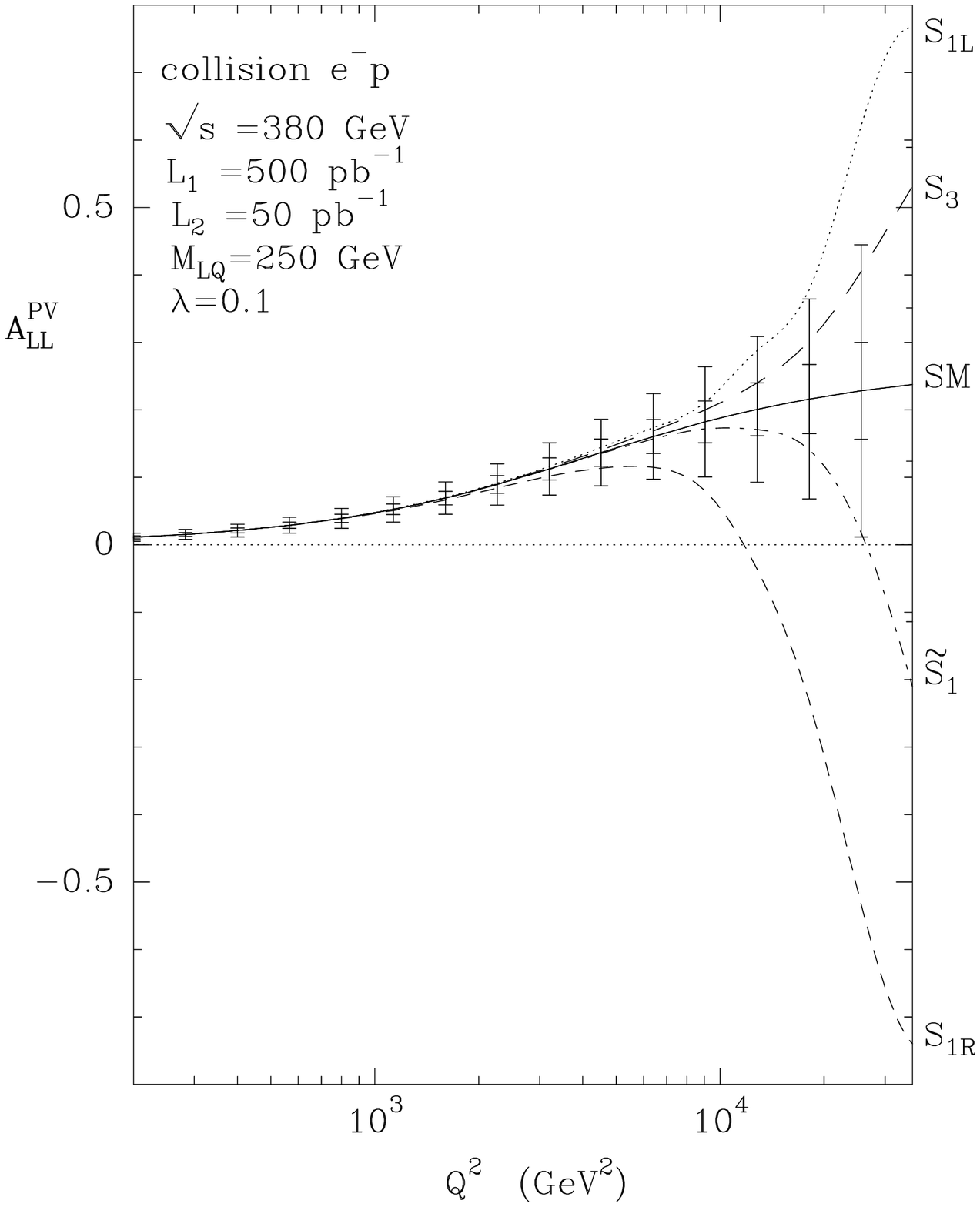,width=6cm}
~ \epsfig{file=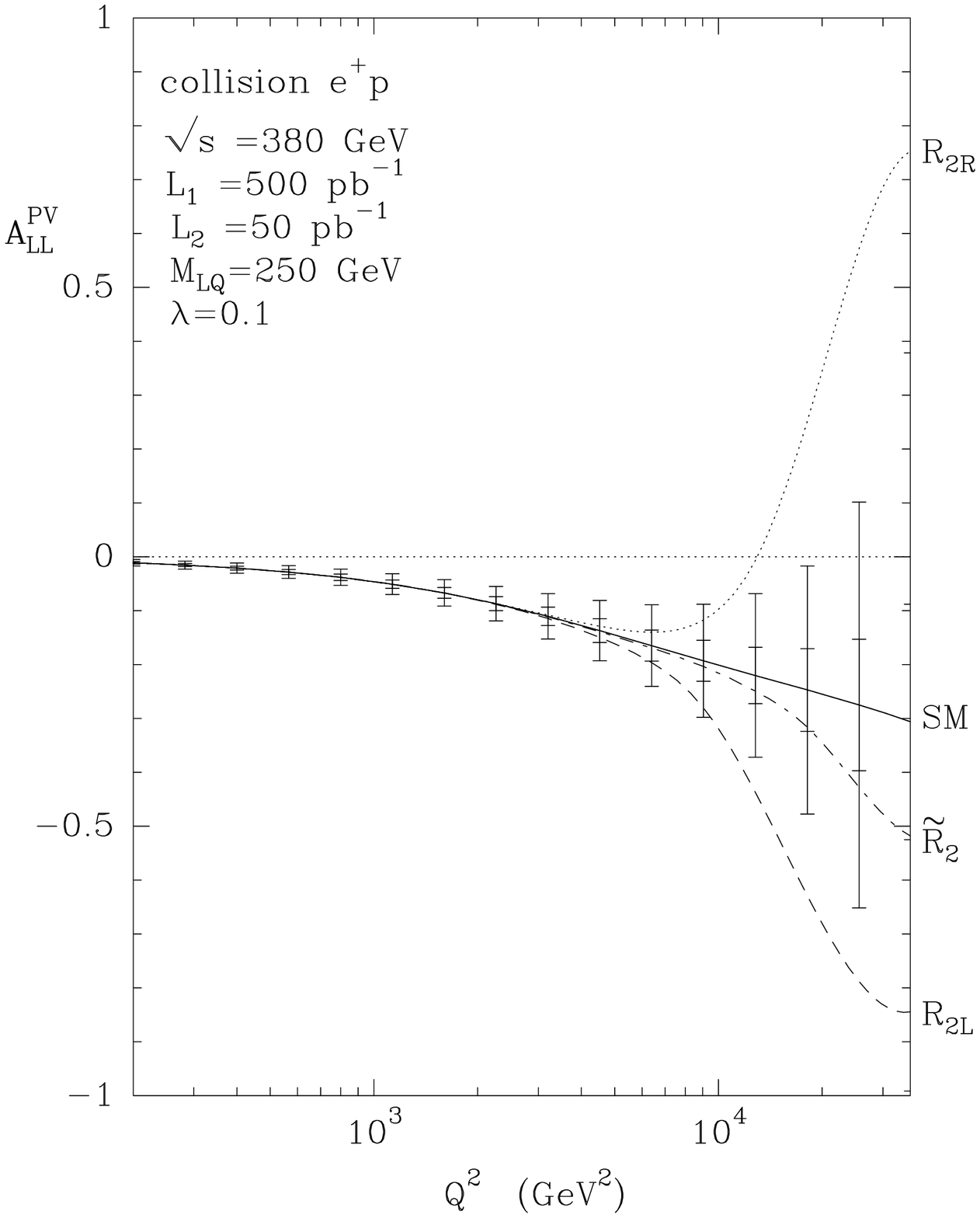,width=6cm}
\end{center}
\caption{Spin asymmetries in leptoquark production. The different
  leptoquark species would yield identical effects in unpolarized
  scattering. See~\cite{virey} for details.}
\label{fig:LQ}
\end{figure}
If the upcoming high luminosity runs at HERA show some evidence for
deviations from the Standard Model, polarization of the proton beam
could help to determine the origin of this deviation. A case study on 
new physics at polarized HERA for leptoquark
production has been reported at this workshop~\cite{virey}. Leptoquarks
are objects carrying both lepton and quark quantum numbers, they can be
created in lepton-quark scattering and would show us as peaks in the
$x$-distribution of deep inelastic scattering events. According to their 
quantum numbers and their coupling structure to leptons and quarks,
leptoquarks can be classified into 
different species~\cite{brw}. HERA could in principle detect 
leptoquarks with a mass up to almost its electron-proton centre-of-mass
energy. This detection is already possible in unpolarized scattering.
However, based on information from unpolarized scattering only, it will
turn out to be very hard to determine the species of the observed
leptoquark. In~\cite{virey}, it was demonstrated that polarization
asymmetries in leptoquark production can be used extract the precise
coupling structure of an observed leptoquark, and thus to determine its
species. Fig.~\ref{fig:LQ} illustrates the potential to discriminate 
different leptoquark species by measuring parity violating spin
asymmetries at $\sqrt{s} = 380$~GeV, which is about the maximum
electron proton centre-of-mass energy that could be eventually reached
at HERA.

\subsection{Fixed target programme at HERA--$\vec{{\rm N}}$}
In addition to the physics programme at the polarized $ep$ collider,
it would be possible to study polarized 
proton-nucleon collisions at a fixed target experiment 
in the polarized HERA proton beam. This proposed 
experiment, HERA--$\vec{{\rm N}}$, would 
require a polarized internal nucleon target and a 
dedicated new spectrometer. It could 
add numerous hadron-hadron observables~\cite{heran}
to the HERA spin physics programme.
Very interesting results are expected here, which are fully
complementary to the RHIC~\cite{bunce} spin physics program. 
The two main issues are probes of the polarized parton distributions in
double spin asymmetries, e.g.~in Drell-Yan process or direct photon
production and investigations into hadronic single spin asymmetries.  

At this workshop, several new investigations for 
single spin asymmetries at HERA--$\vec{{\rm N}}$
were presented: diffractive production of charmed baryons~\cite{morii}
may be used to investigate the spin structure of diffractive
interactions, instanton effects could show up in single spin
asymmetries~\cite{kochelev} and pion production could serve as a probe
of spin transfer in peripheral interactions~\cite{kuraev}.

\section{Conclusions and Outlook}

A variety of new information on the spin structure of the nucleon can be 
expected over the next few years from the continuation of the HERMES
experiment and from COMPASS and RHIC, which are currently under
construction. These experiments will jointly provide the first data 
on the gluon polarization in the nucleon from several complementary 
observables,
and supply new information on the quark contribution to
the nucleon spin. Despite this wealth of new data to be expected, many
questions in spin physics will still remain open. All experiments are
limited in their kinematical reach, in particular for the determination
of the gluon contribution to the proton spin. Moreover, none of these
experiments is sensitive on the structure of the polarized photon -
which is completely unknown at present, but of high importance for
precision predictions of observables to be studied at future
electron-photon or photon-photon colliders.  

Polarization of the HERA proton beam would allow the study of polarized
electron-proton and photon-proton collisions at high energies, which
would allow to measure several aspects of spin structure of proton and
photon which are inaccessible elsewhere. 
Table~\ref{tab:partons} summarizes the information on polarized parton
distributions in the nucleon that can be gained from present and future
experiments. The first block summarizes the current status, the second
block lists the new reactions that can be probed at ongoing and
currently constructed experiments, while the last block illustrates the
improvements that could be made at a polarized HERA collider. 
\begin{table}[t]
\begin{center}
\footnotesize
\begin{tabular}{|l|l|l|}    \hline
{\bf Process/}     &     {\bf Leading order}   & 
{\bf Parton behaviour probed}\\
{\bf Experiment}   &  {\bf subprocess}         &                           \\
\hline
&\hfill \raisebox{-0.5ex}[0.5ex][0.5ex]{}&                      \\ 
{\bf DIS} $\mbox{\boldmath $(\ell N \rightarrow \ell X)$}$ &  $\gamma^*q 
\rightarrow q$ \hfill \hspace*{4pt}& 
 Two structure
functions $\rightarrow$ \\ 
$g^{\ell p}_1,g^{\ell d}_1,g^{\ell n}_1$
& \hfill \hspace*{4pt}& $\sum_q e_q^2 (\Delta q+\Delta \bar{q})$ 
\\ 
(SLAC, EMC/SMC,& \hfill \hspace*{4pt}& 
 $\Delta A_3 = \Delta u+\Delta \bar{u}-\Delta d-\Delta \bar{d}$ \\   
HERMES)& \hfill \hspace*{4pt}& 
\hspace*{1cm}  \\ 
      &\hfill  & \hspace*{1cm}    \\
$\mbox{\boldmath $\ell p, \ell n \rightarrow \ell \pi X$}$    &  
$\gamma^* q \rightarrow q$         
with   &   $\Delta u_V$, $\Delta d_V$, $\Delta \bar{q}$\\
(SMC,HERMES)   & $q = u, d, \bar{u}, \bar{d}$ &     \\[2mm] \hline   
     &     &    \\    
$\mbox{\boldmath $\ell N \rightarrow h^+h^- X$}$ & 
  $\gamma  g\rightarrow  q\bar q$    & $\Delta g$ \\ 
 (HERMES,COMPASS)            &  $\gamma  q\rightarrow  q g$   &
 $(x\approx 0.15, 0.1) $ 
\\[2mm]
     &     &    \\    
$\mbox{\boldmath $\ell N \rightarrow c \bar{c} X$}$ & 
  $\gamma  g\rightarrow  c\bar c$    & $\Delta g $ \\ 
 (COMPASS)            &     &     $(x\approx 0.15)$
\\[2mm]
     &     &    \\    
$\mbox{\boldmath $pp \rightarrow (\gamma^{*},W^{\pm},Z^{0}) X$}$ & 
  $q \bar q \rightarrow \gamma^{*},W^{\pm},Z^{0}$    
& $\Delta u$, $\Delta \bar{u}$, $\Delta d$, $\Delta \bar{d}$ \\ 
 (RHIC)            &     & ($x\gapprox 0.06$)     
\\[2mm]
     &     &    \\    
$\mbox{\boldmath $pp \rightarrow \mbox{{\bf jets }} X$}$ & 
  $q \bar q, qq, qg, gg \rightarrow 2j$    
& $\Delta g$ (?) \\ 
 (RHIC)            &     &     (x\gapprox 0.03)  
\\[2mm]
     &     &    \\    
$\mbox{\boldmath $pp \rightarrow \gamma X$}$ & 
  $qg \rightarrow q\gamma$    
& $\Delta g$ \\ 
 (RHIC)            &  $q\bar q \rightarrow g\gamma$    & (x\gapprox 0.03) 
\\[2mm]\hline 
     &     &    \\    
{\bf DIS} $\mbox{\boldmath $(e^{\pm} p \rightarrow \nu X)$}$ & $W^*q \rightarrow 
q^{\prime}$  \hfill\hspace*{4pt}& Two structure
functions $\rightarrow$ 
\\ 
$g^{\pm}_1,g^{\pm}_5$    &\hfill 
   \hspace*{4pt}& $\Delta u - \Delta \bar{d} - \Delta \bar{s}$
 \\
(HERA)     &\hfill \hspace*{4pt}& $\Delta d + \Delta s - \Delta \bar{u}$
  \\ [2mm]
    &     &    \\   
{\bf DIS (small $x$)}    &   $\gamma^* q \rightarrow q$   &   $\alpha_{q,g}$ 
    \\ 
$g^{ep}_1$ (HERA)   &                  & $(\Delta \bar{q} \sim x^{\alpha_q}, 
 \Delta g \sim  x^{\alpha_g})$    \\[2mm]  
     &     &    \\    
$\mbox{\boldmath $\ell N \rightarrow \ell \mbox{{\bf  ~jets~}} X$}$ & 
  $\gamma^{\star}  g\rightarrow  q\bar q$    & $\Delta g$ \\ 
 (HERA)            &  $\gamma^{\star}  q\rightarrow  q g$   &  $(x\gapprox 0.003) $ 
\\[2mm]
     &     &    \\ 
$\mbox{\boldmath $pp \rightarrow \ell^+ \ell^- X$}$ & 
  $q \bar q \rightarrow \gamma^{\star}$    
& $\Delta \bar{q}$ \\ 
 (HERA-$\vec{{\rm N}}$)            &     & ($x\gapprox 0.15$)     
\\[-6pt]
     &     &    \\    
   \hline

\end{tabular}
\end{center}   
\caption{Information on polarized parton distributions from present and 
future experiments.} 
\label{tab:partons}
\end{table}

At this workshop, new results on the measurability of these key
observables were presented: systematic uncertainties in the 
extraction of the structure function $g_1(x,Q^2)$ from asymmetry
measurements are now better understood, next-to-leading
order QCD corrections are included in 
the extraction of the gluon distribution from 
dijet production and the impact of including charged current HERA data
in global fits has been demonstrated. Concerning the study of polarized
photon structure at HERA, progress has been made with the introduction
of an effective parton approximation, facilitating the extraction of 
parton distributions from measured asymmetries. 

The measurability of several new and potentially interesting
observables has been demonstrated at this workshop: spin asymmetries in 
deep inelastic diffraction could probe the perturbative nature of the
diffractive exchange, asymmetries in deeply virtual compton scattering 
might yield new insights into the proton structure and parity violating
asymmetries could prove to be a valuable tool to study possible effects
of new physics at HERA. 

A new aspect of spin physics at HERA is the possibility of colliding 
the polarized electron beam of a future linear collider with the 
polarized HERA proton beam. First studies on the physics prospects of
this option were presented at the workshop, and work in this direction
is ongoing. Given that the full potential of the linear collider itself
can only be exploited if 
information on the polarized photon structure
is available from other sources, this option highlights the
complementarity of spin physics at HERA with other future projects at
DESY.


\begin{thebibliography}{99}
\bibitem{heraspin} 
J.~Feltesse and A.~Sch\"afer, Proceedings of the workshop
``Future Physics at HERA'', Hamburg 1995/96, eds. G.~Ingelman, A.~De~Roeck 
and R.~Klanner, DESY (Hamburg, 1996), p.757ff. 

\bibitem{spin97}
A.~De Roeck and T.~Gehrmann (eds.), Proceedings of the Workshop
``Physics with Polarized Protons at HERA'', DESY, March - September
1997, DESY--Proceedings 98-001. 

\bibitem{revspin}
A.~Deshpande, these proceedings (hep-ex/9908051).

\bibitem{hermeshad}
HERMES collaboration, K.~Ackerstaff et al., preprint DESY-99-048 
(hep-ex/9906035).

\bibitem{hermesglue}
HERMES collaboration, A.~Airapetian et al., preprint DESY-99-071 
(hep-ex/9907020).

\bibitem{bravar}
A.~Bravar, D.~von Harrach and A.~Kotzinian,  Phys.~Lett. {\bf B421}
(1998) 349. 

\bibitem{teryaev}
O.~Teryaev, these proceedings.

\bibitem{murgia}
M.~Anselmino, M.~Boglione and F.~Murgia, these proceedings (hep-ph/9907269).

\bibitem{efremov}
A.V.~Efremov, O.G.~Smirnova and L.G.~Tkatchev, these proceedings.

\bibitem{meziani}
Z.E.~Meziani, talk given at this workshop.

\bibitem{blumlein}
J.~Bl\"umlein, these proceedings (hep-ph/9907543).

\bibitem{schierholz}
G.~Schierholz et al., these proceedings. 

\bibitem{teryaev2}
J.~Soffer and O.~Teryaev, these proceedings (hep-ph/9906455).

\bibitem{kunne}
F.~Kunne, these proceedings.

\bibitem{bunce}
G.~Bunce, these proceedings.

\bibitem{martin}
O.~Martin and A.~Sch\"afer, these proceedings. 

\bibitem{sowinski}
J.~Sowinski, these proceedings.

\bibitem{morgan}
E.W.N.~Glover and A.G.~Morgan, Z.~Phys.~{\bf C60} (1993) 175.

\bibitem{frixgam}
S.~Frixione, Phys.~Lett.~{\bf B429} (1998) 369.

\bibitem{fv}
S.~Frixione and W.~Vogelsang, preprint hep-ph/9908387. 

\bibitem{chao}
A.~Chao, these proceedings.

\bibitem{deroecketal}
A.~De Roeck et al., Eur.~Phys.~J.~{\bf C6} (1999) 121.

\bibitem{aidala}
C.~Aidala, A.~Deshpande and V.W.~Hughes, these proceedings.

\bibitem{lichtenstadt}
J.~Lichtenstadt, these proceedings.

\bibitem{deshpande}
A.~Deshpande, these proceedings.

\bibitem{smcsmx}
SMC collaboration, B.~Adeva et al., Phys.~Rev.~{\bf D} (in press, 
preprint CERN-EP-99-061).

\bibitem{smallx} 
S.~Bass and A.~De Roeck, these proceedings;\\
M.~Goshtasbpour and P.G.~Ratcliffe, these proceedings;\\
S.M.~Troshin and N.E.~Tyurin, these proceedings (hep-ph/9905564).

\bibitem{badelek}
B.~Badelek, J.~Kiryluk and J.~Kwieci\'{n}ski, preprint hep-ph/9907569.

\bibitem{ber}
J.~Bartels, B.I.~Ermolaev and M.G.~Ryskin, Z. Phys. {\bf C70} (1996) 273; 
{\bf C72} (1996) 627.

\bibitem{ziaja1} 
J.~Kwieci\'{n}ski and B.~Ziaja, these proceedings and preprint
hep-ph/9902440.

\bibitem{ryskin}
J.~Bartels and M.G.~Ryskin, these proceedings.

\bibitem{mana} 
S.~Manaenkov, these proceedings and preprint hep-ph/9903405.

\bibitem{gehrmann}
J.~Bartels, T.~Gehrmann and M.G.~Ryskin, these proceedings and 
Eur.\ Phys.\ J.\ {\bf C} (in press, 
preprint 
hep-ph/9906204).

\bibitem{golo}
S.~Goloskokov, these proceedings (hep-ph/9907429).

\bibitem{dgpast}
G.~R\"adel, A.~De Roeck and M.~Maul in~\cite{spin97}, p.77.

\bibitem{dgnew}
G.~R\"adel and A.~De Roeck, these proceedings.

\bibitem{willfahrt}
E.~Mirkes and S.~Willfahrt in~\cite{spin97}, p.69.

\bibitem{chekanov}
I.~Akushevich and S.~Chekanov, these proceedings. 

\bibitem{ziaja2} 
J.~Kwieci\'{n}ski and B.~Ziaja, these proceedings and preprint 
hep-ph/9906499.

\bibitem{phoprodold}
J.M.~Butterworth, N.~Goodman, M.~Stratmann and W.~Vogelsang
in~\cite{spin97}, p.120. 

\bibitem{frixione}
D.~de Florian and S.~Frixione, Phys.~Lett.~{\bf B457} (1999) 236.

\bibitem{maxwell}
B.L.~Combridge and C.J.~Maxwell, Phys.~Lett.~{\bf B239} (1984) 429.

\bibitem{stratmann}
M.~Stratmann and W.~Vogelsang, these proceedings (hep-ph/9907470). 

\bibitem{sudoh}
T.~Morii and K.~Sudoh, these proceedings.

\bibitem{arestov}
Yu.~Arestov, these proceedings.

\bibitem{dvcs}
X.~Ji, Phys.~Rev.~{\bf D55} (1997) 7114;\\
A.V.~Radyushkin, Phys.~Rev.~{\bf D56} (1997) 5524. 

\bibitem{strikman}
A.~Freund and M.~Strikman, these proceedings.

\bibitem{virey}
A.~De Roeck, P.~Taxil, E.~Tugcu and J.M.~Virey, these proceedings.

\bibitem{brw}
W.~Buchm\"uller, R.~R\"uckl and D.~Wyler, Phys.~Lett.~{\bf B191} (1987) 442.

\bibitem{heran}
V.A.~Korotkov and W.D.~Nowak, these proceedings (hep-ph/9908490).

\bibitem{morii}
N.~Kochelev, T.~Morii and S.~Oyama, these proceedings.

\bibitem{kochelev}
N.~Kochelev, these proceedings (hep-ph/9905497).

\bibitem{kuraev}
A.~Akhmedov, I.~Akushevich, E.~Kuraev and P.~Ratcliffe, these
proceedings and preprint hep-ph/9902418.

\end{thebibliography}
\end{document}